\documentclass[]{spie}  %>>> use for US letter paper
\usepackage[utf8]{inputenc}
\usepackage[T1]{fontenc}
\usepackage{lmodern}
\usepackage{textcomp}

\usepackage{aas_macros}
\usepackage{siunitx}

\usepackage{amsmath,amsfonts,amssymb}
\usepackage{graphicx}
\usepackage[colorlinks=true, allcolors=blue]{hyperref}

\title{A next generation upgraded observing platform for the automated Birmingham Solar Oscillations Network (BiSON)}

\author[a,b]{S. J. Hale}
\author[a,b]{W. J. Chaplin}
\author[a,b]{G. R. Davies}
\author[a,b]{Y. P. Elsworth}
\affil[a]{School of Physics and Astronomy, University of Birmingham, Edgbaston, Birmingham B15 2TT, United Kingdom}
\affil[b]{Stellar Astrophysics Centre, Department of Physics and Astronomy, Aarhus University, Ny Munkegade 120, DK-8000 Aarhus C, Denmark}

\authorinfo{Further author information:\\Send correspondence to S. J. Hale \textless s.j.hale@bham.ac.uk\textgreater}

\begin{document} 
\maketitle

%%%%%%%%%%%%%%%%%%%%%%%%%%%%%%%%%%%%%%%%%%%%%%%%%%%%%%%%%%%%%%%%%%%%%%%%%%%%%%%%

\begin{abstract}
The Birmingham Solar Oscillations Network (BiSON) is a collection of
ground-based automated telescopes observing oscillations of the Sun.
The network has been operating since the early 1990s.  We present
development work on a prototype next generation observation platform,
BiSON:NG, based almost entirely on inexpensive off-the-shelf
components, and where the footprint is reduced to a size that can be
inexpensively installed on the roof of an existing
building. Continuous development is essential in ensuring that
automated networks such as BiSON are well placed to observe the next
solar cycle and beyond.
\end{abstract}

% Include a list of keywords after the abstract 

%%%%%%%%%%%%%%%%%%%%%%%%%%%%%%%%%%%%%%%%%%%%%%%%%%%%%%%%%%%%%%%%%%%%%%%%%%%%%%%%

% -*- coding: utf-8 -*-
%
% INTRODUCTION.TEX
%
%   Steven Hale
%   2020 March 27
%   Birmingham, UK
%
% Astronomical Telescopes + Instrumentation 2020 in Yokohama, Japan
%

\section{INTRODUCTION}
\label{sec:intro}  % \label{} allows reference to this section

\begin{figure}[ht]
  \begin{center}
    \begin{tabular}{c} %% tabular useful for creating an array of images 
      \includegraphics[width=0.8\textwidth]{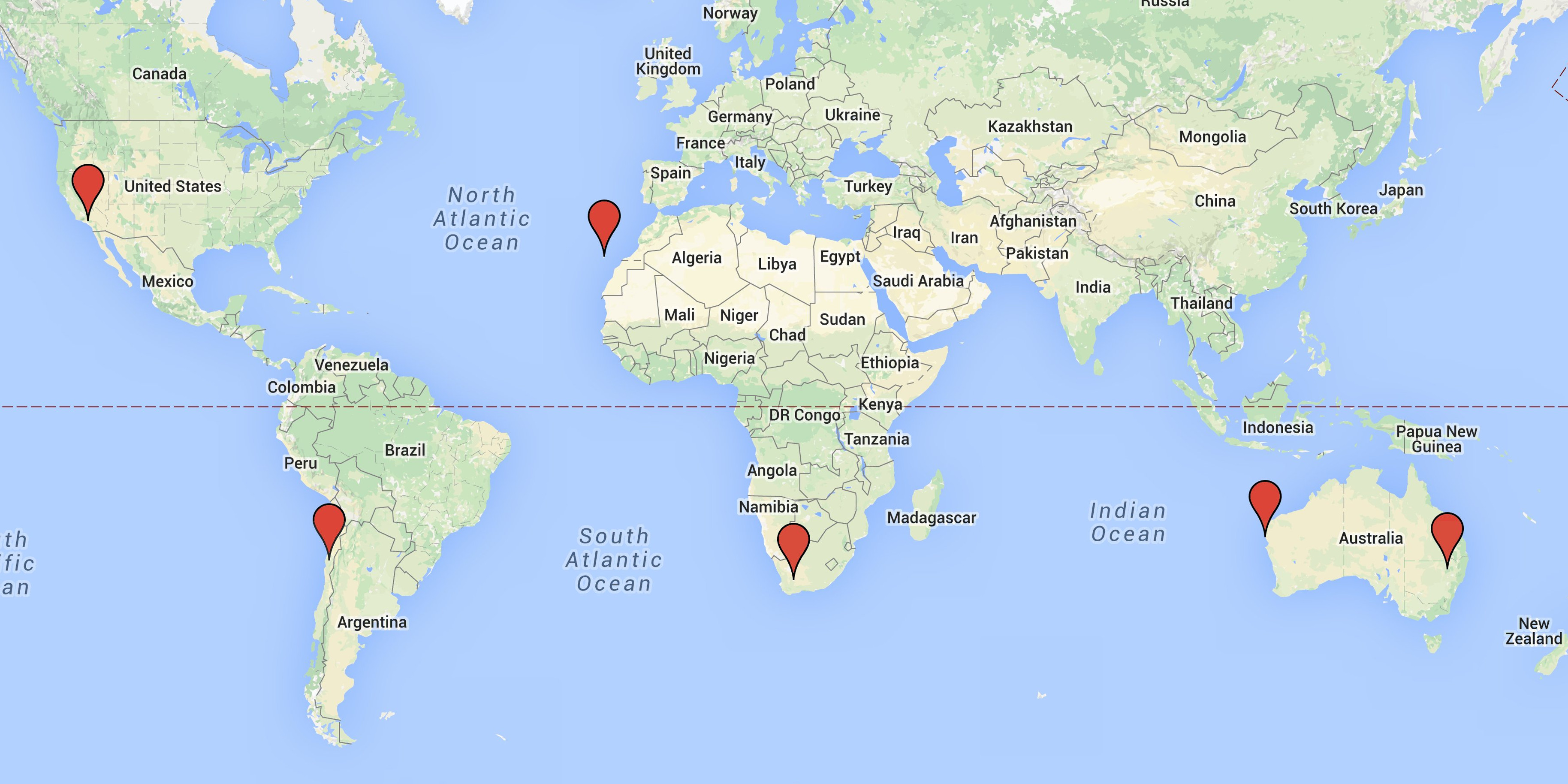}
    \end{tabular}
  \end{center}
  \caption{\label{fig:network_map} The six station Birmingham Solar
    Oscillations Network (BiSON). Image credit: Google Maps}
\end{figure} 

The Birmingham Solar Oscillations Network (BiSON) is a collection of
ground-based automated solar telescopes observing helioseismic
oscillations of the Sun-as-a-star.  The network as it stands today was
completed in~1992, and consists of six remote sites shown in
Fig.~\ref{fig:network_map}.  A detailed history of the network is
available in Ref.~\citenum{2016SoPh..291....1H} and~\citenum{halephd}.
Here, we present development work on a prototype next generation small
observation platform, BiSON:NG, a significantly miniaturised system
based almost entirely on inexpensive consumer-grade off-the-shelf
components and commodity hardware.

The existing BiSON solar observatory infrastructure requires a 4~metre
observatory dome housing a large equatorial mount.  By making use of
modern fibre optics it is now possible to split the instrumentation
from the front end collection optics, and so allow use of a much
smaller mount with considerably lower payload capacity, and
subsequently a smaller telescope enclosure.  This can be combined with
taking advantage of many years of electronic miniaturisation, which
has produced micro-controllers and single-board computers allowing old
single-purpose hardware to be retired.  The resulting physical
footprint of BiSON:NG is reduced to a size that can be easily and
inexpensively installed on the roof of an existing building, requiring
only a small enclosure rather than a dedicated observatory, with no
reduction in performance.  In the following sections we will discuss
some of the key difficulties that have been overcome to allow full
automation of consumer-grade astronomy hardware.

%%%%%%%%%%%%%%%%%%%%%%%%%%%%%%%%%%%%%%%%%%%%%%%%%%%%%%%%%%%%%%%%%%%%%%%%%%%%%%%%

% -*- coding: utf-8 -*-
%
% MOUNT.TEX
%
%   Steven Hale
%   2020 March 27
%   Birmingham, UK
%
% Astronomical Telescopes + Instrumentation 2020 in Yokohama, Japan
%

\section{AUTOMATION OF A CONSUMER-GRADE TELESCOPE MOUNT}
\label{sec:mems}  % \label{} allows reference to this section

There are two key issues when attempting to automate a small
commercial~off-the-shelf (COTS) telescope mount.  The first is access
to a communications protocol with a published application programming
interface~(API) in order to allow full computer control of the mount.
Many COTS mounts do not have facility for computer control at all, and
rely only on the supplied proprietary control handset.  In cases where
the handset offers an ASCOM compatible control interface this
typically does not work without first calibrating the mount position
within the handset software, and they offer no facility to store a
permanent calibration since they are expected to be packed away at the
end of each observing session.  Typically movement is tracked by
counting stepper-motor steps once initial alignment has been
completed, and so the calibration is lost as soon as the mount is
moved by hand or the power is interrupted.  Larger, more expensive,
research-grade mounts employ absolute position encoders to avoid this
problem.  A fully automated telescope must be able to recover after a
power failure without manual recalibration.

The communication and control problem can be solved by simply choosing
only manufacturers that open-source and publish their control API.
The issue of alignment and recovery from power failure without user
intervention is rather more tricky.  Potentially a mount could be
retro-fitted with absolute position encoders, or limit switches to
indicate a home position, but this moves away from the intention of
simple off-the-shelf use.  Instead, it is possible to make use of
microelectromechanical systems (MEMS) sensors to directly detect the
attitude of the telescope.  Such devices are most commonly used in
smartphones and tablets, with accelerometers used to control screen
orientation, and magnetometers used to determine heading when
navigating.  These inexpensive sensors can be easily mounted on the
telescope itself and avoid costly modifications to the mount.

\begin{figure}[t]
  \begin{center}
    \begin{tabular}{c c} %% tabular useful for creating an array of images 
      \includegraphics[height=0.35\textwidth]{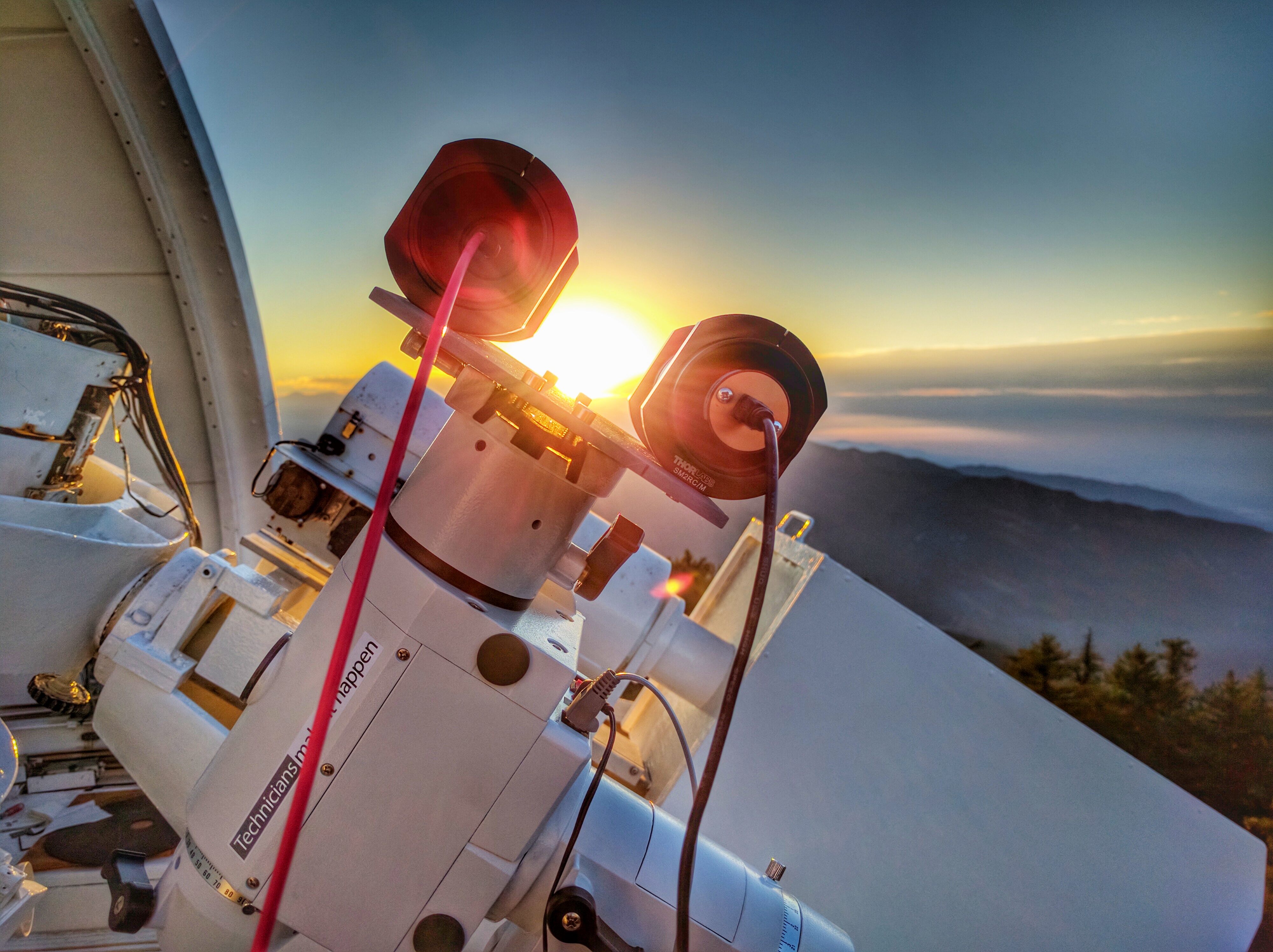} &
      \includegraphics[height=0.35\textwidth]{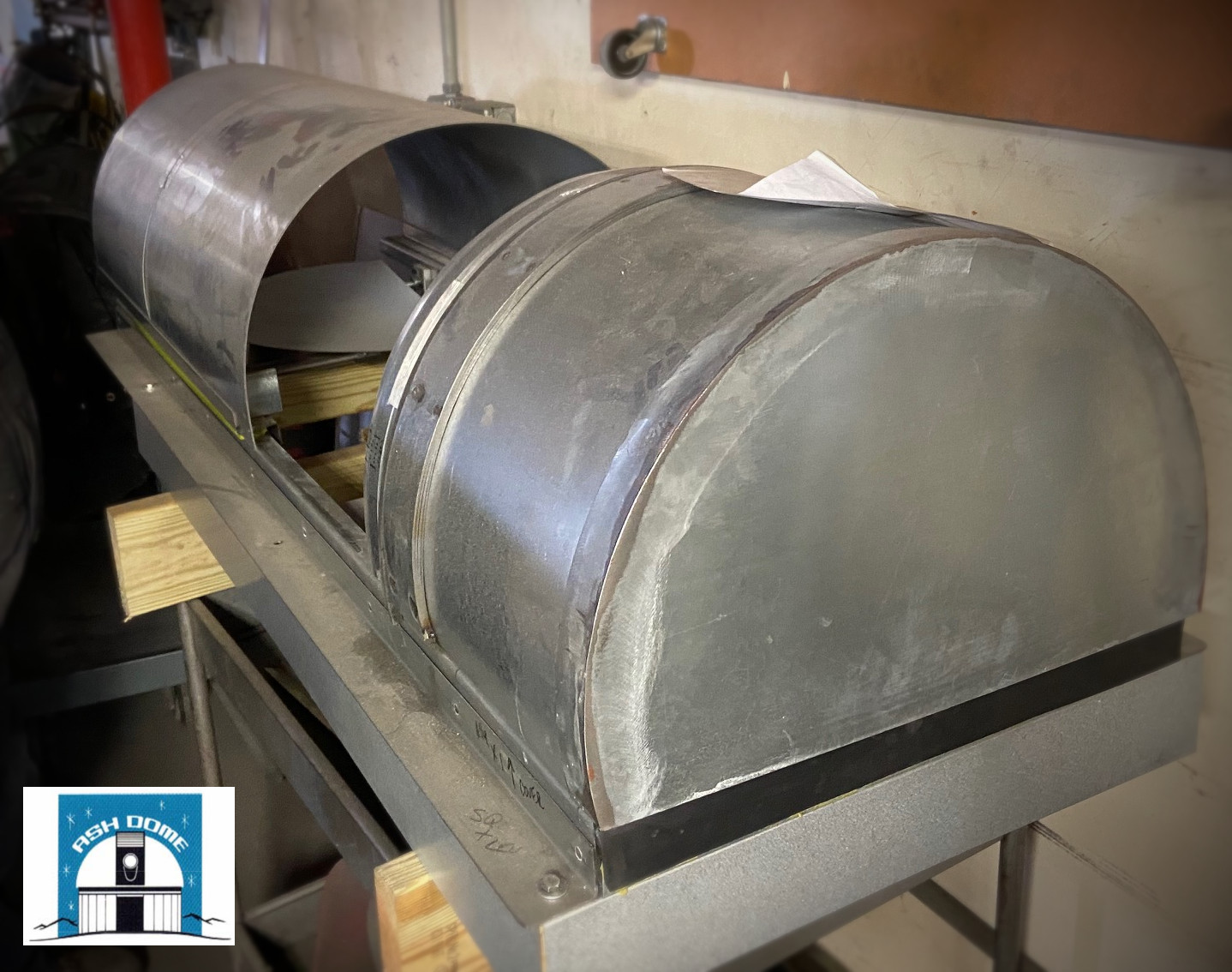}
    \end{tabular}
  \end{center}
  \caption{\label{fig:enclosure} Left: Sunrise at the Mount Wilson
    (Hale) Observatory\cite{epapers2987}. A small consumer-grade
    equatorial mount was configured with a binocular arrangement of a
    CCD camera for guiding, and optical fibre collection optics for
    data acquisition trial. Photo credit: S.\,J.\,Hale. Right: A small
    telescope enclosure and weather-proofing prototype with rolling
    roof.  Photo credit: Ash Manufacturing Company\cite{ashdome}.}
\end{figure}

A consumer-grade equatorial mount was trialled at the Mount~Wilson
(Hale) Observatory 60~foot solar tower, an existing BiSON site, in
order to determine the possible accuracy of MEMS attitude sensors, and
the precision of solar autoguiding\cite{bison382}.
Figure~\ref{fig:enclosure}, left panel, shows the temporary binocular
configuration of fibre collection optics for solar data acquisition,
and a solar-filtered CCD camera for guiding.
Figure~\ref{fig:enclosure}, right panel, shows a prototype automated
weather-proof housing with a rolling-roof suitable for a small mount
and optics.  The final design and construction of an enclosure is
subject to further work.

In subsection~\ref{sec:mems-acc} below we demonstrate determining
telescope attitude from an ADXL345 three-axis accelerometer, in
subsection~\ref{sec:mems-mag} acquiring magnetic heading from a
HMC5883L three-axis magnetometer, and in subsection~\ref{sec:guiding}
guiding the mount on the centroid of the solar disc estimated using
computer vision techniques.

%%%%%%%%%%%%%%%%%%%%%%%%%%%%%%%%%%%%%%%%%%%%%%%%%%%%%%%%%%%%%%%%%%%%%%%%%%%%%%%%

\subsection{Determining attitude using a three-axis accelerometer}
\label{sec:mems-acc}

A MEMS accelerometer measures linear acceleration, and this includes
the acceleration due to the Earth's gravitational field vector.
Inside the device, a micro proof-mass is suspended by restoring
springs, and deflection of the proof-mass due to acceleration is
detected by measuring changes in capacitance between the proof-mass
and sensing plates.  An embedded micro-controller digitises the signal
and allows the data to be output over a digital serial bus.  A total
of three measurement axes are required in order to sense acceleration
in all directions.  A thorough discussion of calibration of a
three-axis accelerometer and extracting pitch and roll angles is given
in Ref.~\citenum{an3461} and~\citenum{an4399}.  We will now summarise
the calibration process and go on to apply the techniques to an
Analog~Devices~ADXL345 accelerometer fixed to the optical tube of an
equatorially mounted telescope.

The Analog~Devices~ADXL345 accelerometer is factory calibrated to
output values in units of $g$, with a precision of~\num{4}\,milli-$g$
per least-significant bit and a dynamic range of~\num{\pm16}\,$g$.
The accelerometer measures both the gravitational field vector, and
linear acceleration due to motion.  Measurement accuracy of the
gravitational field is reduced when the device is subject to
additional external accelerations, however on a quasi-static design
such as a telescope mount this is not a concern.  The factory
calibration is a 6-term model providing gain and offset parameters for
each of the three channels, and this generally provides adequate
results in the typical use case of orienting a smartphone.
Re-calibrating after the device is installed allows for greater
precision to be achieved by correcting for thermal stresses introduced
during soldering, and the convenience of allowing installation at an
arbitrary angle -- the device axes do not need to align with the
system axes.  A general 12-parameter calibrated output
$\mathbf{G}_{12}$ can be defined in terms of the factory calibration
$\mathbf{G}_{f}$ by~\cite[eq.~35]{an4399},
\begin{equation}\label{eq:G12_1}
    \mathbf{G}_{12}
     = \begin{pmatrix} G_{12x} \\ G_{12y} \\ G_{12z}\end{pmatrix}
     = \mathbf{W}\mathbf{G}_{f} + \mathbf{V}
     = \begin{pmatrix} W_{xx} & W_{xy} & W_{xz}\\
                       W_{yx} & W_{yy} & W_{yz}\\
                       W_{zx} & W_{zy} & W_{zz}\end{pmatrix}
       \begin{pmatrix} G_{fx} \\ G_{fy} \\ G_{fz}\end{pmatrix} +
       \begin{pmatrix} V_{x} \\ V_{y} \\ V_{z}\end{pmatrix} ~,
\end{equation}
where the gain matrix $\mathbf{W}$ includes any rotation of the
integrated circuit package and also corrects for all possible
cross-talk interactions, and $\mathbf{V}$ are the channel offsets.

If $M$ measurements are used for the 12-parameter calibration, then
the $i$-th measurement at pitch angle $\theta[i]$ and roll angle
$\phi[i]$ becomes~\cite[eq.~36]{an4399},
\begin{equation}%\label{eq:G12_2}
     \begin{pmatrix} G_{12x}[i] \\ G_{12y}[i] \\ G_{12z}[i]\end{pmatrix}
     = \begin{pmatrix} W_{xx} & W_{xy} & W_{xz}\\
                       W_{yx} & W_{yy} & W_{yz}\\
                       W_{zx} & W_{zy} & W_{zz}\end{pmatrix}
       \begin{pmatrix} G_{fx}[i] \\ G_{fy}[i] \\ G_{fz}[i]\end{pmatrix} +
       \begin{pmatrix} V_{x} \\ V_{y} \\ V_{z}\end{pmatrix}
     = \begin{pmatrix} -\sin(\theta[i]) \\ \cos(\theta[i]) \sin(\phi[i])\\ \cos(\theta[i]) \cos(\phi[i])\end{pmatrix} ~,
\end{equation}
which can be decomposed into three equations,
\begin{equation}%\label{eq:G12_3}
\begin{split}
   W_{xx} G_{fx}[i] + W_{xy} G_{fy}[i] + W_{xz} G_{fz}[i] + V_x &= -\sin(\theta[i]) \\
   W_{yx} G_{fx}[i] + W_{yy} G_{fy}[i] + W_{yz} G_{fz}[i] + V_y &=  \cos(\theta[i]) \sin(\phi[i]) \\
   W_{zx} G_{fx}[i] + W_{zy} G_{fy}[i] + W_{zz} G_{fz}[i] + V_z &=  \cos(\theta[i]) \cos(\phi[i]) ~,
\end{split}
\end{equation}
with residuals,
\begin{equation}\label{eq:G12_4}
\begin{split}
   r_x[i] &= -\sin(\theta[i]) - W_{xx} G_{fx}[i] - W_{xy} G_{fy}[i] - W_{xz} G_{fz}[i] - V_x \\
   r_y[i] &=  \cos(\theta[i]) \sin(\phi[i]) - W_{yx} G_{fx}[i] - W_{yy} G_{fy}[i] - W_{yz} G_{fz}[i] - V_y \\
   r_z[i] &=  \cos(\theta[i]) \cos(\phi[i]) - W_{zx} G_{fx}[i] - W_{zy} G_{fy}[i] - W_{zz} G_{fz}[i] - V_z ~.
\end{split}
\end{equation}

If we now consider only the $x$-component, equation~\ref{eq:G12_4} can
be simplified to~\cite[eq.~41]{an4399},
\begin{equation}%\label{eq:G12_5}
   r_x = Y_x - \mathbf{X}\beta_x ~,
\end{equation}
where $r_x$ is the $M$-length array of residuals to the calibration
fit, $Y_x$ is the array of $x$-components of the gravitation field for
the true measured angle, $X$ is the matrix of accelerometer
measurements, and finally $\beta_x$ is the solution vector for four of
the calibration parameters.  The optimum least squares solution for
$\beta_x$ can be found by making use of the Normal Equations for least
squares optimisation,
\begin{equation}\label{eq:leastsquares}
    \beta = (\mathbf{X}^T \mathbf{X})^{-1} \mathbf{X}^T \mathbf{Y} ~,
\end{equation}
such that~\cite[eq.~46]{an4399},
\begin{equation}%\label{eq:G12_6}
    \beta_x = \begin{pmatrix} W_{xx} \\ W_{xy} \\ W_{xz} \\ V_x \end{pmatrix}
            = (\mathbf{X}^T \mathbf{X})^{-1} \mathbf{X}^T Y_x
            = (\mathbf{X}^T \mathbf{X})^{-1} \mathbf{X}^T
              \begin{pmatrix} -\sin{\theta[0]} \\  -\sin{\theta[1]} \\ \dots \\ -\sin{\theta[M-1]} \end{pmatrix} ~,
\end{equation}
and similarly for the remaining two axes.  In order to solve for
12-parameters the minimum number of measurements $M$ is four, and
these should be well distributed in the measurement space.  More
measurements give a more robust calibration.  The calibrated
accelerations are subsequently determined using
equation~\ref{eq:G12_1}.

\begin{figure}[t]
  \begin{center}
    \begin{tabular}{c c} %% tabular useful for creating an array of images 
      \includegraphics[width=0.45\textwidth]{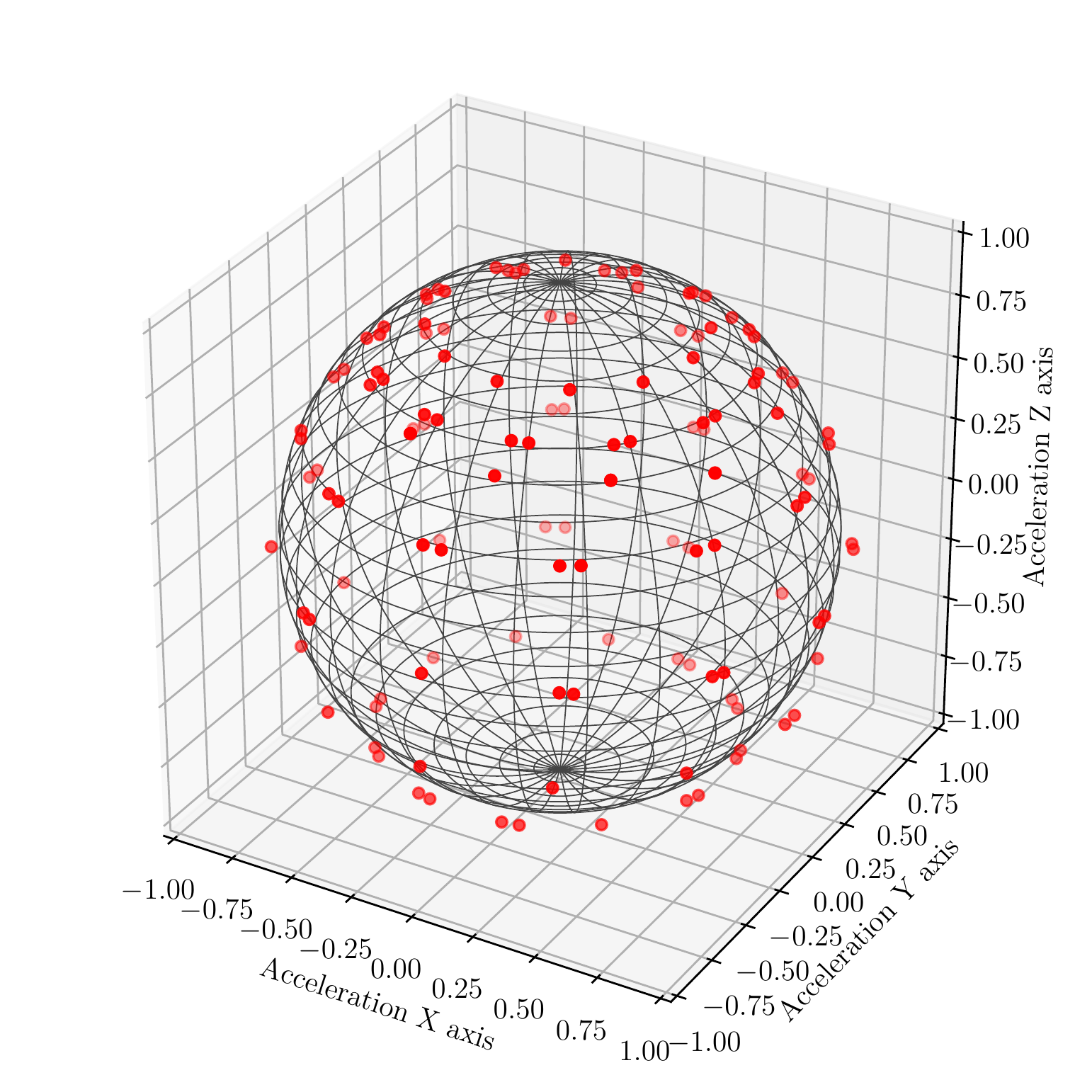} &
      \includegraphics[width=0.45\textwidth]{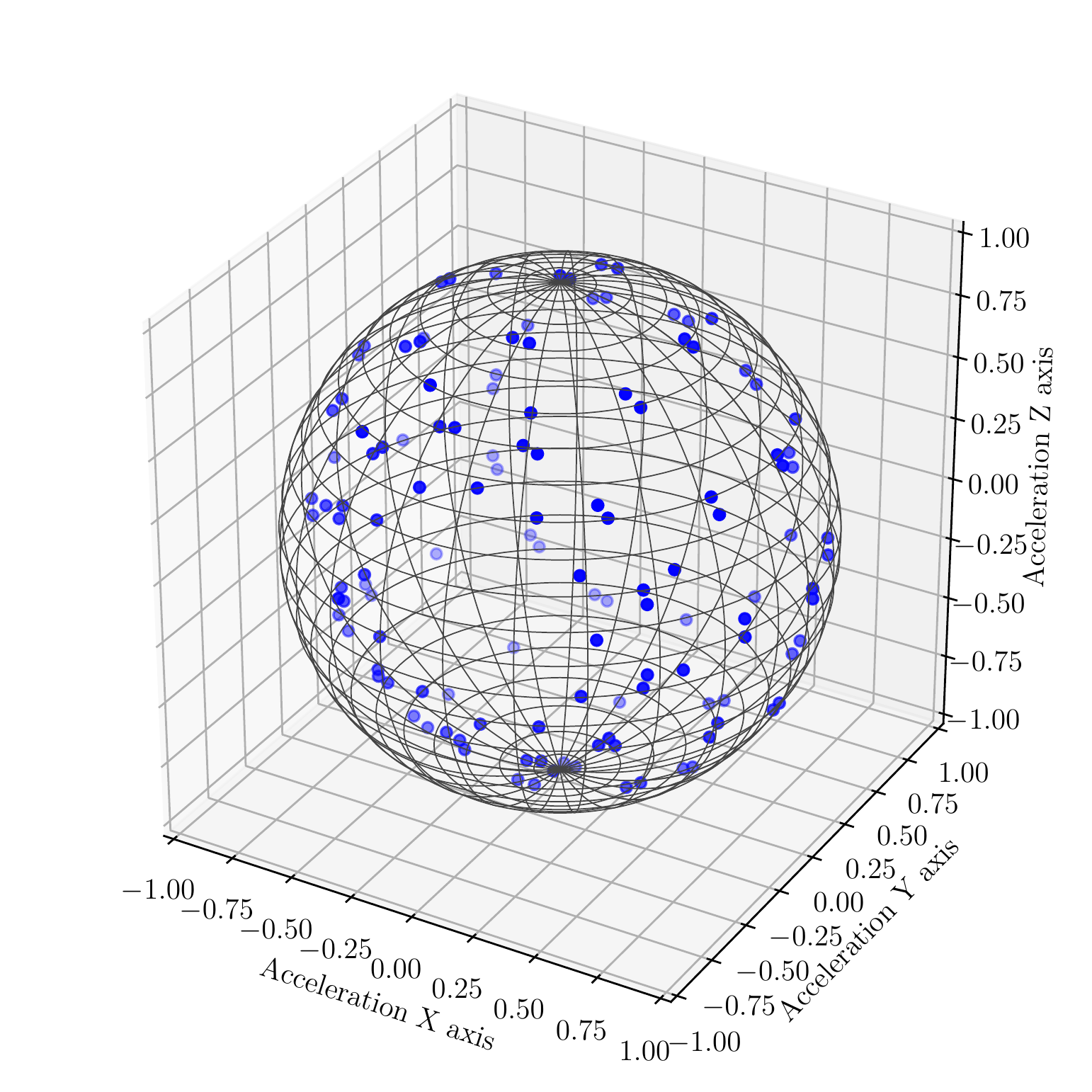}
    \end{tabular}
  \end{center}
  \caption{\label{fig:mems_acc}{Data from an Analog~Devices~ADXL345
      accelerometer fixed to the optical tube of an equatorially
      mounted telescope, slewed through its full range of motion.
      Left: The raw data with standard factory calibration.  Right:
      The data after 12-parameter calibration.  All points now sit on
      the surface of a sphere of radius $1\,g$.}}
\end{figure}

Figure~\ref{fig:mems_acc} shows a range of measurements made whilst
the mount was slewed slowly through its full range of motion.  Both
factory calibration and improved 12-parameter calibration are shown.
The roll and pitch angles can be computed from the calibrated
measurement matrix $\mathbf{G}_{12}$ using~\cite[eq.~25--26]{an3461},
\begin{equation}\label{eq:acc_roll1}
    \phi_{xyz} = \tan^{-1}\left(\frac{G_{12y}}{G_{12z}}\right) ~,
\end{equation}
and,
\begin{equation}%\label{eq:acc_pitch}
    \theta_{xyz} = \tan^{-1}\left(\frac{-G_{12x}}{\sqrt{{G_{12y}}^2 + {G_{12z}}^2}}\right) ~,
\end{equation}
where $\phi$ and $\theta$ are the measured roll and pitch angles
respectively, and the subscript $xyz$ notes that the angles are
according to the aerospace rotation sequence $\mathbf{R}_{xyz}$ where
rotation is first in yaw, then pitch, then roll.

Equation~\ref{eq:acc_roll1} becomes unstable when the telescope is
pointing near the zenith, since the $x$-axis becomes aligned with the
gravitational field vector and enters a condition known as Gimbal
Lock.  Any rotation in roll can no longer be detected and roll becomes
undefined.  A common work-around for this problem when using aerospace
rotation sequence $\mathbf{R}_{xyz}$ is to modify
equation~\ref{eq:acc_roll1} to include in the denominator a fraction
$\mu$ of the $x$-axis measurement whilst remembering to maintain the
sign of $G_{12z}$ after taking the square root~\cite[eq.~38]{an3461},
\begin{equation}%\label{eq:acc_roll2}
    \phi_{xyz} = \tan^{-1}\left(\frac{G_{12y}}{\pm \sqrt{{G_{12z}}^2 + \mu{G_{12x}}^2}}\right) ~,
\end{equation}
such that $\phi$ is slowly driven to zero as the telescope approaches
a vertical orientation.  An additional ambiguity is caused while the
telescope is vertical, since with an equatorial mount it becomes
impossible to determine if the telescope is on the east or west side
of the pier.  This ambiguity can be resolved by the addition of a
second accelerometer installed directly on the polar axis of the
mount.

In testing, the accelerometer achieved an accuracy of
approximately~\SI{\pm4}{\degree} in pitch and~\SI{\pm6}{\degree} in
roll.  The reduced performance in roll is due to the above work-around
at high pitch angles.  Performance from both axes is similar if only
moderate pitch angles are considered.  {M{\'e}sz{\'a}ros}
et.\,al.\,trialled the two-accelerometer technique by mounting two
Freescale MMA8453Q accelerometers on their telescope at Konkoly
Observatory located at the Piszk{\'e}stet{\"o} Mountain
station\cite{2014PASP..126..769M}, and achieved accuracy to better
than a degree with more sophisticated calibration, by making use of
full temperature compensation, and taking good care of power supply
stability via a custom printed circuit board.  The simpler treatment
shown here is adequate for a solar autoguider with a capture angle of
a few degrees.

Accelerometers are capable of completely resolving telescope attitude
only with an equatorial telescope mount.  With an altitude-azimuth
mount, only changes in altitude (pitch) can be measured.  Rotations in
azimuth (yaw) are aligned with the gravitational field vector and
cannot be detected, and so a different sensor is required.  Next, we
look at using a three-axis magnetometer to detect the orientation of
Earth's magnetic field and directly measure the telescope ``heading''
angle.

%%%%%%%%%%%%%%%%%%%%%%%%%%%%%%%%%%%%%%%%%%%%%%%%%%%%%%%%%%%%%%%%%%%%%%%%%%%%%%%%

\subsection{Determining heading using a three-axis magnetometer}
\label{sec:mems-mag}

In order to detect rotation about a vector parallel to Earth's
gravitational field vector (i.e., yaw, heading, azimuth) we need to be
able to detect rotation within Earth's magnetic field, otherwise known
as an electronic compass.  We used a Honeywell HMC5883L three-axis
magnetometer, which uses magneto-resistive sensors to measure both the
direction and the magnitude of Earth's magnetic field.  The Honeywell
HMC5883L magnetometer is factory calibrated to output values in units
of~\si{\micro\tesla}, with a dynamic range of~\SI{800}{\micro\tesla},
and an embedded 12-bit ADC providing a
documented~\SIrange{1}{2}{\degree} heading accuracy.  As with the
accelerometer, a magnetometer requires a final in-situ calibration in
order to achieve the best precision.  A thorough discussion of
calibration of a three-axis magnetometer is given in
Refs.~\citenum{an4246, an4247, an4248, an4249}.  We will now summarise
the calibration process and go on to apply the techniques to the
Honeywell HMC5883L, including tilt-compensation based on earlier
results from the Analog~Devices~ADXL345 accelerometer.

A magnetic vector measured at the device $\mathbf{B}_{d}$ can be
defined after arbitrary device rotation in terms of the local
geomagnetic field vector $\mathbf{B}$ by~\cite[eq.~5]{an4246},
\begin{equation}\label{eq:B4_1}
    \mathbf{B}_{d}
     = \mathbf{W} \mathbf{R}_x(\phi) \mathbf{R}_y(\theta) \mathbf{R}_z(\psi)
     \lVert \mathbf{B} \rVert \begin{pmatrix} \cos{\delta} \\ 0 \\ \sin{\delta}\end{pmatrix}
     + \mathbf{V} ~,
\end{equation}
where $\phi$, $\theta$, and $\psi$ are roll, pitch, and yaw angles as
previously, $\delta$ is the magnetic inclination at the measurement
location, $\mathbf{V}$ is the ``hard-iron'' offset vector, and
$\mathbf{W}$ is the ``soft-iron'' gain matrix.  So-called hard-iron
offsets are magnetic fields generated by nearby permanent magnets,
such as other components on the PCB, and motors in the telescope
mount.  These components are generally in fixed positions and rotate
with the device, and so they appear as an additive magnetic field
vector within the reference frame of the magnetometer.  So-called
soft-iron interference is due to temporary induction of magnetic
fields in otherwise normally unmagnetised components, such as the
sheet steel of the telescope mount and housing, caused by the
geomagnetic field itself.  Soft-iron effects are much more complicated
to model since they depend on the orientation of the device within the
geomagnetic field, and typically suffer magnetic hysteresis effects as
the device rotates.  The soft-iron matrix $\mathbf{W}$ is a 9-element
matrix similar to that used during calibration of the accelerometer in
equation~\ref{eq:G12_1}, and in addition to calibrating soft-iron
effects also calibrates rotation of the integrated circuit package and
corrects for all possible cross-talk interactions and gain variations
between channels.  In addition to removing offsets, the measurements
from a magnetometer need to be derotated back to the flat plane where
$\phi = \theta = 0$, since as with a typical analogue compass it works
only when held level.  Rearranging equation~\ref{eq:B4_1} for $\psi$
we get~\cite[eq.~6]{an4246},
\begin{equation}%\label{eq:B4_2}
\begin{split}
   \mathbf{R}_z(\psi) \lVert \mathbf{B} \rVert \begin{pmatrix} \cos{\delta} \\ 0 \\ \sin{\delta}\end{pmatrix}
   &= \begin{pmatrix}  \cos{\psi} & \sin{\psi} & 0 \\
                      -\sin{\psi} & \cos(\psi) & 0 \\
                                0 &          0 & 1 \end{pmatrix}
     \lVert \mathbf{B} \rVert \begin{pmatrix} \cos{\delta} \\ 0 \\ \sin{\delta}\end{pmatrix} \\
   &= \mathbf{R}_y(-\theta) \mathbf{R}_x(-\phi) \mathbf{W}^{-1}
     (\mathbf{B}_{d} - \mathbf{V}) ~,
\end{split}
\end{equation}
and so~\cite[eq.~9]{an4246},
\begin{equation}%\label{eq:B4_3}
   \begin{pmatrix} B_{fx} \\ B_{fy} \\ B_{fz}\end{pmatrix}
   = \begin{pmatrix}  \cos{\psi} \cos{\delta} \lVert \mathbf{B} \rVert \\
                     -\sin{\psi} \cos{\delta} \lVert \mathbf{B} \rVert \\
                                 \sin{\delta} \lVert \mathbf{B} \rVert \end{pmatrix}
   = \mathbf{R}_y(-\theta) \mathbf{R}_x(-\phi) \mathbf{W}^{-1}
     (\mathbf{B}_{d} - \mathbf{V}) ~,
\end{equation}
where $\mathbf{B}_{f}$ is equal to the magnetometer measurements with
both soft-iron and hard-iron effects removed, and derotated to a flat
plane where the $z$-component $B_{fz}$ is equal to $\lVert \mathbf{B}
\rVert \sin{\delta}$.  The yaw angle, or compass heading, is found
from~\cite[eq.~10]{an4246},
\begin{equation}\label{eq:mag_yaw}
    \psi = \tan^{-1}\left(\frac{-B_{fy}}{B_{fx}}\right) ~,
\end{equation}
and this simply requires the addition of the known local magnetic
declination to convert from magnetic north to true north, and so
obtain the azimuthal angle.

In many cases the soft-iron effects are insignificant, and only the
hard-iron offsets dominate.  This allows a simplification during
calibration since only four parameters need to be determined, and
these are the magnitude of the geomagnetic field strength
$\lVert \mathbf{B} \rVert$, and the three components of the hard-iron
vector $\mathbf{V}$.  The soft-iron matrix $\mathbf{W}$ becomes the
identity matrix.  With these assumptions we can follow the same
calibration procedure as for the accelerometer, by developing a
performance function to be minimised by optimising the calibration fit
and again using equation~\ref{eq:leastsquares} to solve the fit
through matrix algebra.

\begin{figure}[ht]
  \begin{center}
    \begin{tabular}{c c} %% tabular useful for creating an array of images 
      \includegraphics[width=0.45\textwidth]{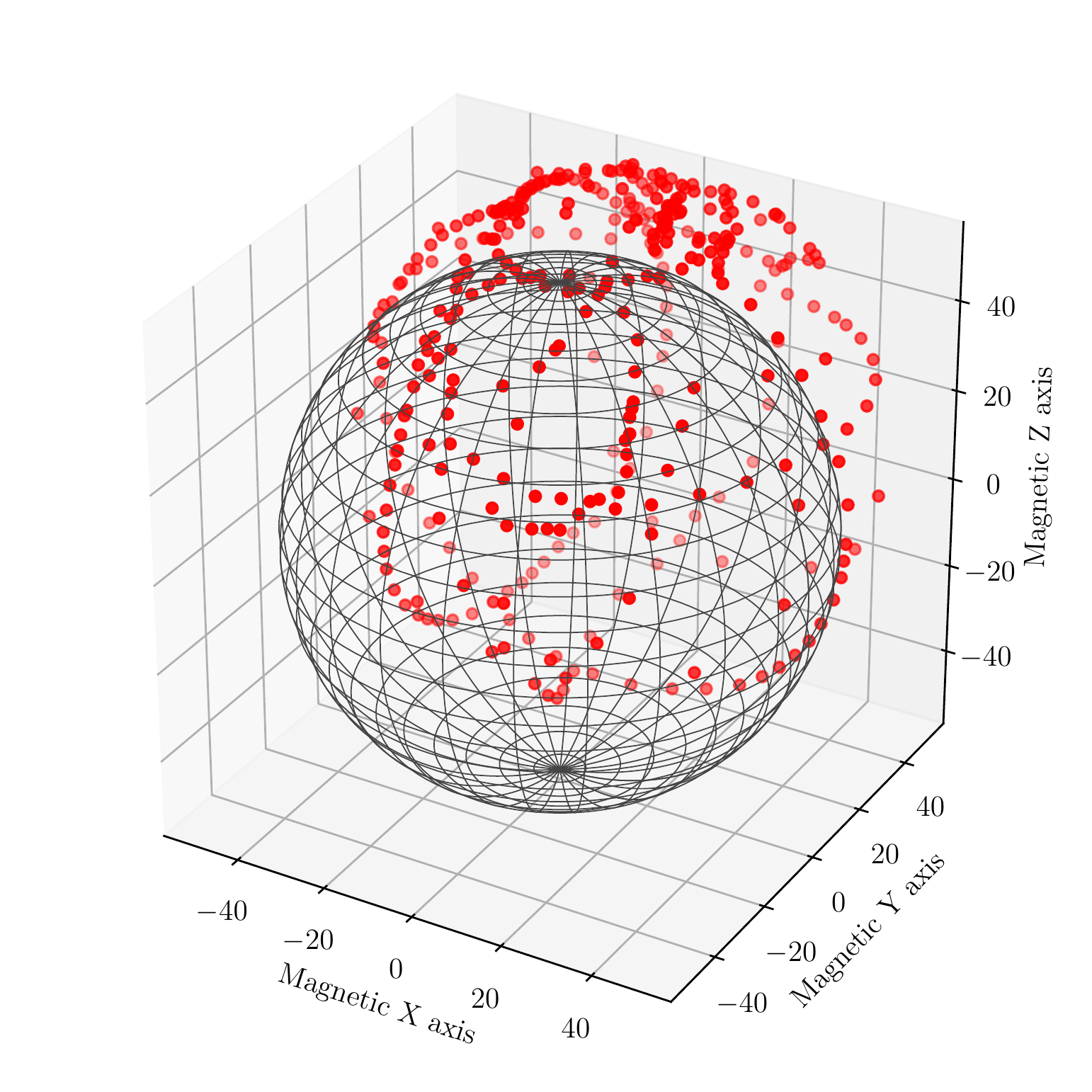} &
      \includegraphics[width=0.45\textwidth]{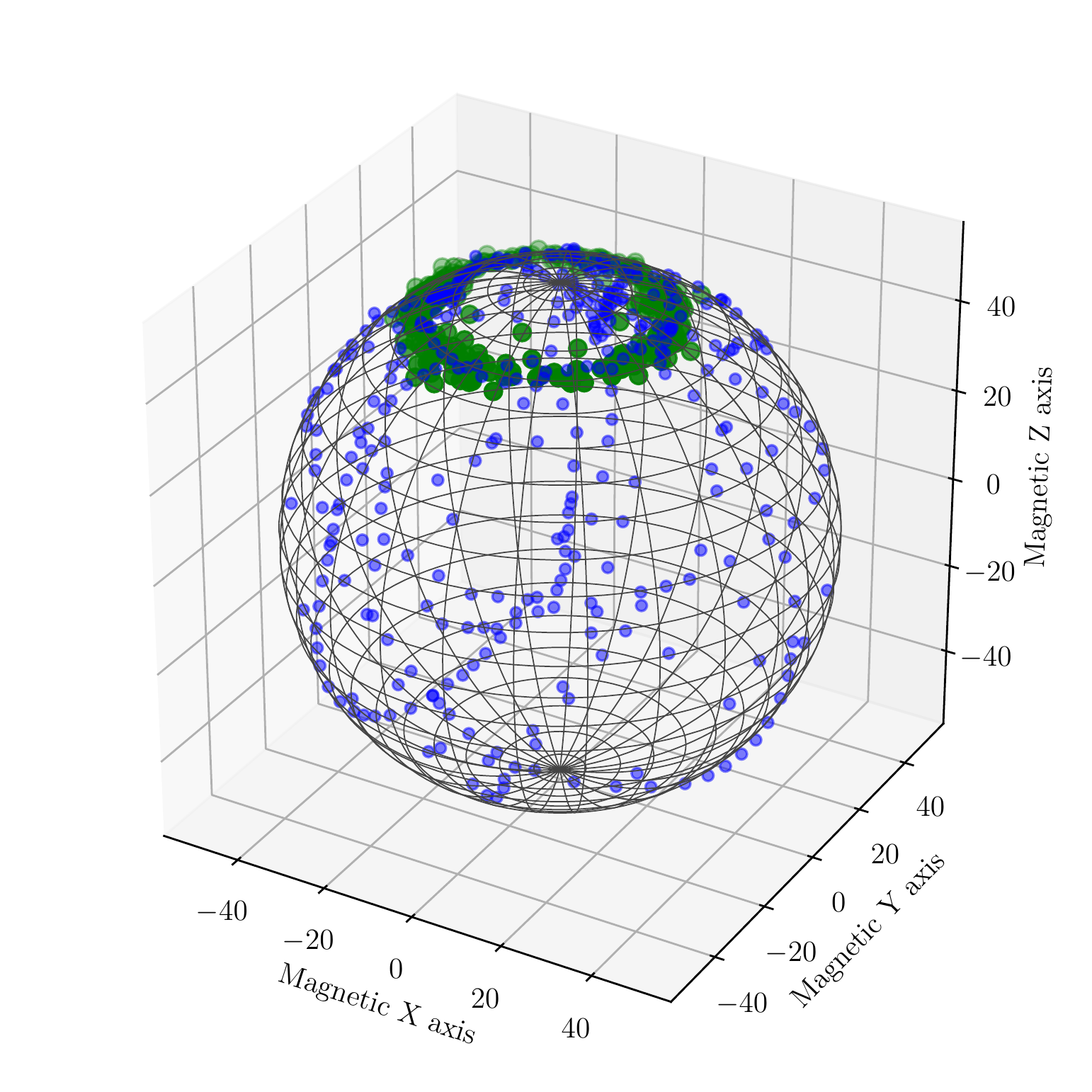}
    \end{tabular}
  \end{center}
  \caption{\label{fig:mems_mag} {Data from a Honeywell HMC5883L
      magnetometer slewed randomly through the full range of motion.
      Left: The raw data with a hard-iron offset. Right: The blue dots
      show the data after 4-parameter calibration removing the
      hard-iron offset.  All points now sit on the surface of a sphere
      centred at the origin, with radius equal to the local
      geomagnetic field strength in Birmingham calibrated at
      \SI{54.5}{\micro\tesla}.  The green dots show the same data
      after derotation to the flat plane where $\phi = \theta = 0$ and
      the $z$-component $B_{fz}$ is equal to $\lVert \mathbf{B} \rVert
      \sin{\delta}$.}}
\end{figure}

Figure~\ref{fig:mems_mag} shows a range of measurements made by a
Honeywell HMC5883L magnetometer slewed randomly through the full range
of motion.  It is clear from the left panel of
Figure~\ref{fig:mems_mag} that the data do indeed suffer from a
hard-iron offset, which moves the data away from the origin.  After
applying the 4-parameter calibration, shown by the blue dots in the
right panel of Figure~\ref{fig:mems_mag}, the data now sit on the
surface of a sphere centred at the origin, with radius equal to the
local geomagnetic field strength in Birmingham calibrated
at~\SI{54.5}{\micro\tesla}.  Our assumption that the soft-iron effects
are insignificant is shown to be true, since these would have the
effect of distorting the data away from a perfect sphere and into an
ellipse.  The green dots show the same data after derotation, by
making use of the pitch and roll angles determined by the
accelerometer, to the flat plane where $\phi = \theta = 0$ and the
$z$-component $B_{fz}$ is equal to $\lVert \mathbf{B} \rVert
\sin{\delta}$.  These data are now calibrated and ready for the
heading to be extracted using equation~\ref{eq:mag_yaw}.  The
resulting accuracy after calibration is
approximately~\SI{\pm5}{\degree}.  With further work and more careful
calibration, it is expected to achieve the manufacturers stated
performance of~\SIrange{1}{2}{\degree} resolution.  With this simple
calibration, and as with the accelerometer, the performance is
adequate for a solar autoguider with a capture angle of a few degrees.

Having shown that it is possible to obtain sufficiently accurate
absolute pointing information of an inexpensive telescope mount at
power-on, we will now go on to consider fine guiding of a mount once
the target has been acquired by the coarse pointing.

%%%%%%%%%%%%%%%%%%%%%%%%%%%%%%%%%%%%%%%%%%%%%%%%%%%%%%%%%%%%%%%%%%%%%%%%%%%%%%%%

\subsection{Solar auto-guiding using computer vision}
\label{sec:guiding}

After the Sun is brought within the capture angle of the autoguider
camera by the coarse MEMS-based pointing, mount control is handled
purely via imaged-based guiding much like any other telescope.  The
camera trialled has a~\SI{6.4}{\milli\metre}
by~\SI{4.75}{\milli\metre} CCD with~\SI{4.65}{\micro\metre} square
pixels in a~\num{1392} by \num{1040}~array.  When coupled with
an~\SI{80}{\milli\metre} focal length objective lens this produces an
approximate field of view of~\SI{4.6}{\degree} by~\SI{3.5}{\degree},
where each pixel has about~\SI{12}{\arcsecond} field of view.  The Sun
has an extent of about~\SI{32}{\arcminute} and so produces an image
about 160~pixels in diameter on the sensor.  Two filters were used to
bring the image within the dynamic range of the CCD chip.  These were
a neutral density filter with optical density of~\num{5} (i.e,
transmission of approximately $10^{-5}$), and a~\SI{10}{\nano\metre}
bandpass filter centered on~\SI{780}{\nano\metre}. Images were read
from the camera as frequently as possible, approximately once per
second.  The image exposure time was~\SI{79}{\milli\second}, but the
cadence was restricted by the CCD read-out time and USB transfer rate.

Images from the camera were processed to determine the position of the
Sun to sub-pixel accuracy using the OpenCV (Open-source
Computer-Vision) library\cite{opencv_library}, running on an
inexpensive single-board-computer. The centroid position of the solar
disc was found using by applying a black and white threshold, finding
the contours in the image, and subsequently reading out the contour
centroid position.  The guiding error was determined by comparing the
current solar centroid position with a desired target value.  The
position error for each axis was then passed through a
proportional–integral–derivative (PID) control loop feedback mechanism
in order to determine the correct mount drive rate, and the mount
motors updated with the new drive rate using the motor API published
by the mount manufacturer.  The PID control algorithm is a servo
feedback system that can be defined simply by,
\begin{equation}
  \mathrm{Output} = K_P \, e(t) +
                    K_I \int{e(t)\,dt} +
                    K_D \frac{\mathrm{d}e(t)}{\mathrm{d}{t}} ~,
\end{equation}
where $e$ is the control error defined as the desired setpoint minus
the current value, and $K_P$, $K_I$, and $K_D$ are the proportional,
integral, and derivative coefficients.  The solar guiding algorithm
made use of only the proportional and integral parameters.

\begin{figure}[t]
  \begin{center}
    \begin{tabular}{c} %% tabular useful for creating an array of images 
      \includegraphics[scale=1.0]{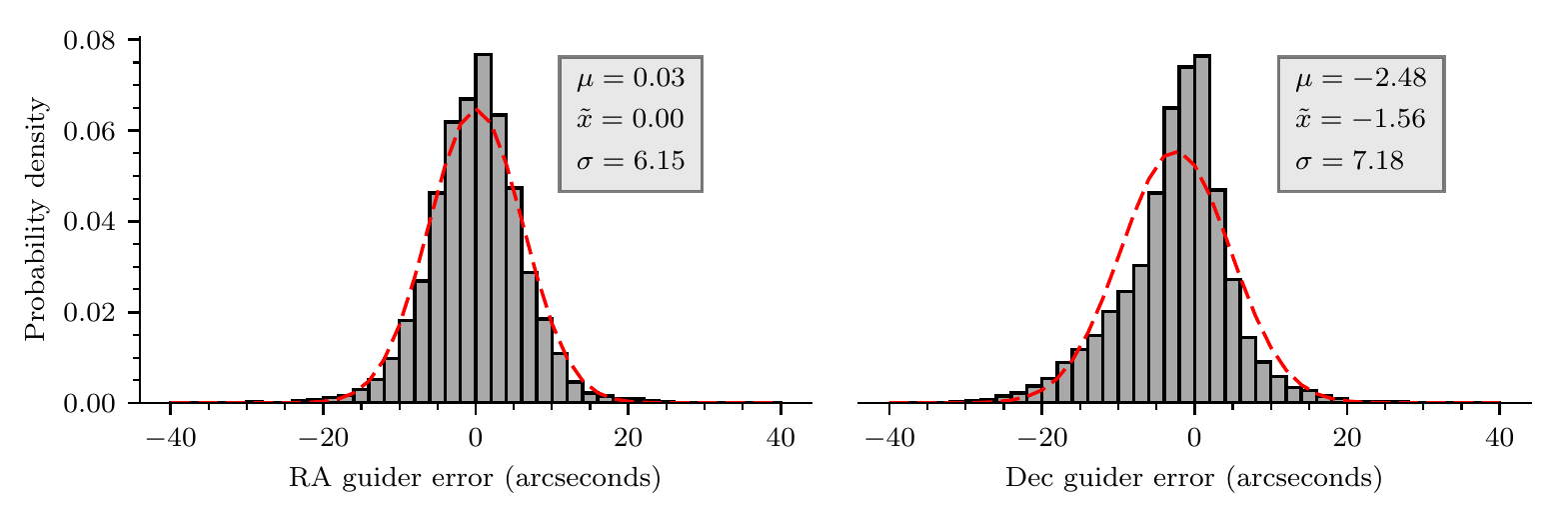}
    \end{tabular}
  \end{center}
  \caption{\label{fig:mo160925_hist} Guider performance histogram.
    The dashed red line indicates the equivalent Gaussian for the
    measured median and standard deviation.}
\end{figure} 

The distribution of position-error for both axes, calibrated in
arcseconds, is shown in Figure~\ref{fig:mo160925_hist}.  The control
algorithm achieves performance on both axes to better
than~\SI{\pm8}{\arcsecond} when logged at an
approximate~\SI{1}{\hertz} cadence over an entire day.  This is more
than sufficient for Sun-as-a-star observations when considering the
Sun is over~\SI{1800}{\arcsecond} in observed diameter.

By making use of these techniques is it possible to achieve full
automation of consumer-grade telescope mounts.  In the next section we
will discuss prototype updates to the BiSON control systems.

%%%%%%%%%%%%%%%%%%%%%%%%%%%%%%%%%%%%%%%%%%%%%%%%%%%%%%%%%%%%%%%%%%%%%%%%%%%%%%%%

% -*- coding: utf-8 -*-
%
% ACQUISITION.TEX
%
%   Steven Hale
%   2020 March 27
%   Birmingham, UK
%
% Astronomical Telescopes + Instrumentation 2020 in Yokohama, Japan
%

\section{SUPERVISORY CONTROL AND DATA ACQUISITION}
\label{sec:acquisition}

\begin{figure}[t]
  \begin{center}
    \begin{tabular}{c} %% tabular useful for creating an array of images 
      \includegraphics[width=0.7\textwidth]{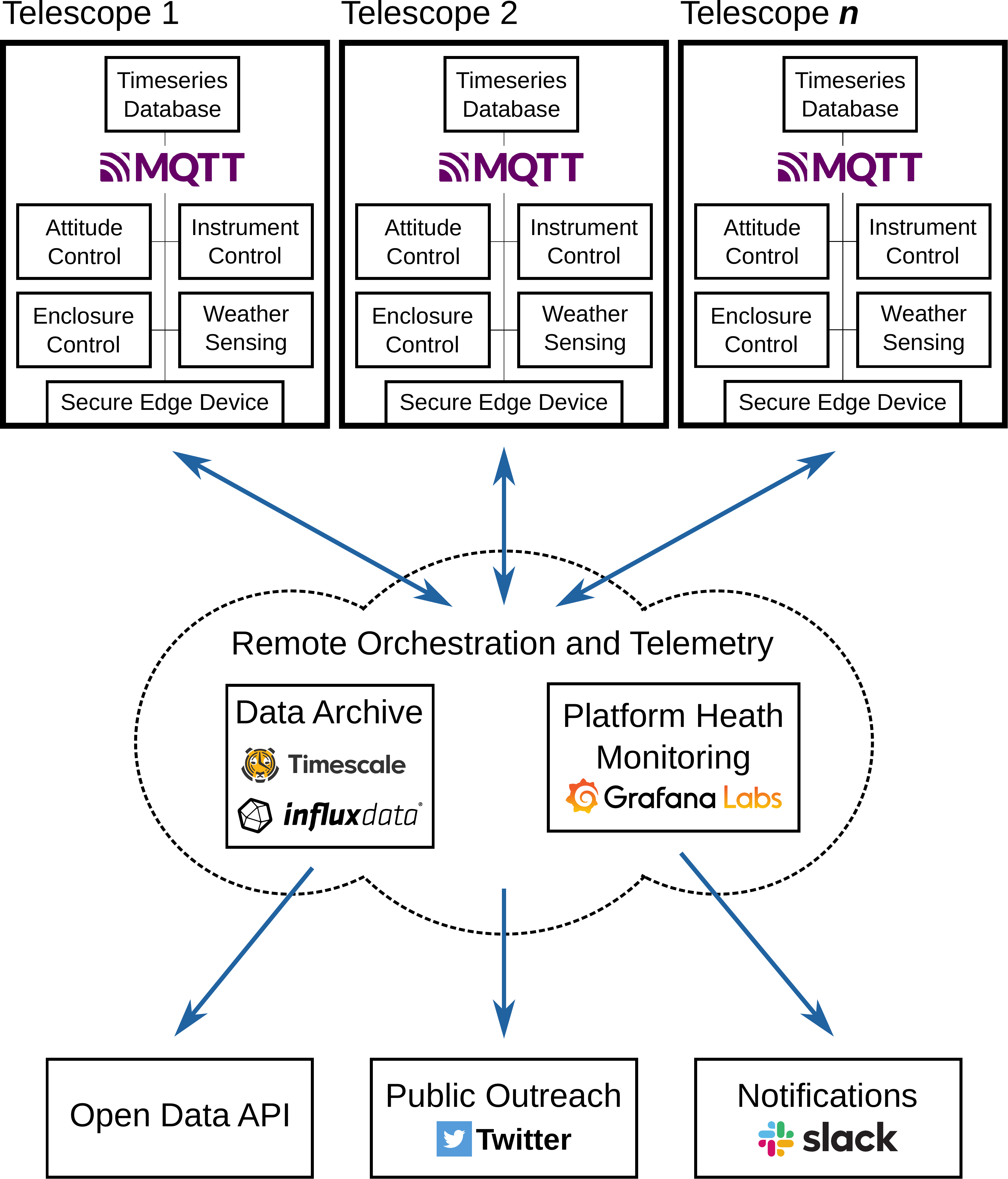}
      %\rule{.1pt}{5cm} \rule{10cm}{.1pt} \rule{.1pt}{5cm}
    \end{tabular}
  \end{center}
  \caption{\label{fig:microservices} A block diagram of the proposed
    microservice architecture for BiSON:NG, that is easily scalable to
    many small telescopes in a globally distributed network.}
\end{figure} 

In the early 1990s, the original network control system was based
around a standard desktop PC running Microsoft~DOS, and a Keithley
System~570 digital input/output interface for connection to dome,
telescope, and instrumentation control hardware.  The computer acted
as a centralised controller, handling all systems from dome and mount
pointing, to temperature stabilisation and data readout, using a
Supervisory Control and Data Acquisition (SCADA) monolithic control
system architecture.  The data would be retrieved each day over a
long-distance telephone call via dial-up modem, and later over the
internet.  The Keithley interface devices were used until 2002, when
the DOS PCs were replaced with the current systems running GNU/Linux
with hardware interconnects predominately via RS-232, although a PCI
general purpose digital-IO card is still used to interface to some
legacy equipment.

The so-called ``Industry~4.0'', the fourth industrial revolution, has
since brought about inter-connected machines, devices, and sensors --
the ``Internet of Things'' (IoT) -- that have allowed control systems
to become decentralised.  By migrating from a monolithic to a
microservice based architecture (MSA) in a Distributed Control System
(DCS) it is possible to further reduce both the physical footprint,
complexity, and deployment cost.  A microservice follows the Unix
philosophy of ``Do One Thing and Do It Well'', where each service is a
single purpose process capable of handling a request independently.
Such decoupled processes allow a modular approach to system design,
reducing development into more manageable components and removes the
need to understand the entire monolithic system.  Independent modules
also mean they can be independently upgraded or replaced, removing the
common risk of being forced to redesign an entire system simply
because one component has become obsolete.  When all devices and
sensors are connected to every other service, it can also become
possible trigger maintenance processes autonomously by monitoring data
from all points in the overall system and predicting potential
failures.  Such flexibility is essential in the operation of a
worldwide constellation of small telescopes, that inevitably become
heterogeneous due to gradual roll-out of repairs and upgrades.
Figure~\ref{fig:microservices} shows a block diagram of the proposed
MSA for BiSON:NG, and also for implementation at the existing network
telescopes.

Communication between microservices is typically handled using
lightweight protocols with well-defined inputs and outputs.  Most IoT
devices communicate over internet protocol networks making use of
industry standard Ethernet or WiFi.  This has a significant advantage
in allowing many devices to be easily connected using standard network
switches, rather than relying on the limited connectivity options
available on a single PC.  Components such as sensors and controllers
from different manufacturers need to communicate using standardised
communication protocols.  As is common with standards there are
several from which to choose, such as OPC~UA (Open Platform
Communications Unified Architecture), DDS (Data Distribution Service),
MQTT (Message Queuing Telemetry Transport), CoAP (Constrained
Application Protocol), Extensible Messaging and Presence Protocol
(XMPP), Advanced Message Queuing Protocol (AMQP), and
others\cite{8755050}.  Whilst there is no one best protocol, we have
chosen to use MQTT due to it being designed for low-bandwidth
connections from remote locations.  It is extremely lightweight, with
open source examples of implementation on micro-controllers where a
small code footprint is essential, and single-board computers, both of
which we use extensively.

MQTT is an ISO~standard client/server publish/subscribe protocol
developed by IBM\cite{mqtt}.  The protocol requires a message broker
through which clients communicate, with messages organised into
topics.  Messages are broadcast on a one-to-many distribution basis.
The protocol can scale easily up to hundreds or even thousands of
devices, allowing decoupling of applications since any service with an
interest in a particular data feed can simply subscribe to the
relevant topic.  Several quality-of-service (QoS) options are
available, ensuring messages are delivered either at most once, at
least once, or exactly once, providing robust communication even over
high-latency or unreliable networks.  Typically only a single broker
is required, although multiple brokers can be linked for either
redundancy, load sharing, or connecting between local and remote
sites.  There are also security advantages to using MQTT, since
communication can be easily authenticated over a Transport Layer
Security (TLS) encrypted channel.  The broker manages all security
credentials and certificates, and also tracks the client connection
states allowing rapid notification of a disconnected service.

A number of free and open-source packages are used where possible.
Data archival is handled both locally at each telescope, in order to
survive network outages, and remotely for aggregation.  A
timeseries-specific database is used, such as
Timescale\cite{timescale} and InfluxDB\cite{influxdb}, to ensure
performance at scale.  Platform monitoring and telemetry visualisation
is produced by Grafana\cite{grafana}, with system log messages written
locally and, where necessary, published to a Slack\cite{slack} channel
for urgent notification.  Public outreach is possible through
automatic status updates via social media such as
Twitter\cite{twitter}.

Through the upgrades discussed here, we move from a full rack of
electronics to physically small services closely coupled near their
respective areas of instrumentation.

%%%%%%%%%%%%%%%%%%%%%%%%%%%%%%%%%%%%%%%%%%%%%%%%%%%%%%%%%%%%%%%%%%%%%%%%%%%%%%%%

% -*- coding: utf-8 -*-
%
% CONCLUSION.TEX
%
%   Steven Hale
%   2020 March 27
%   Birmingham, UK
%
% Astronomical Telescopes + Instrumentation 2020 in Yokohama, Japan
%

\section{CONCLUSION}
\label{sec:conclusion}

The Birmingham Solar Oscillations Network has achieved an average
annual duty cycle of around 82\% since commissioning in 1992,
providing an unparalleled baseline of unresolved-Sun helioseismic
observations\cite{2016SoPh..291....1H}.  The aim of observing
potential solar gravity-modes ($g$-modes) requires much lower noise
levels over long time periods than currently achieved by any
Sun-as-a-star observations, since they are expected to have very low
amplitudes and low frequencies\cite{2010A&ARv..18..197A}.  The
instrumental noise level is dominated by atmospheric
scintillation\cite{2020PASP..132c4501H} and that of solar origin, but
both can be beaten down by combining multiple incoherent measurements
from many simultaneous
observations\cite{2014MNRAS.441.3009D,2017MNRAS.472.3256L}, and so
there is a need to considerably increase the number of BiSON observing
sites.

We have shown here that it is possible to achieve full automation of
an inexpensive consumer-grade telescope mount through the addition of
MEMS inertial sensors, within a small physical package requiring only
a basic weather-proof automated enclosure.  A microservices control
architecture allows the control systems to run entirely on inexpensive
single-board computers and micro-controllers, removing the need for a
full rack of electronics and again reducing cost.  Initial trials of
these small form factor techniques at three existing BiSON sites have
shown an improvement in performance through the reduction of certain
noise sources such as guider
errors\cite{halephd,bison384,bison385,bison389,bison390}.  In order to
ensure that BiSON is well placed to observe the next solar cycle and
beyond, continuous development and improvement is essential.

The impact of BiSON:NG extends far wider than the field of solar
astronomy.  A large network of small inexpensive robotic telescopes
can easily be made dual purpose, for both solar and stellar astronomy.
Telescopes as small as 6~inches have research applications where,
perhaps operated by schools and Citizen Scientists, they are capable
of follow-up observations of exoplanet
transits\cite{Dragomir_2020,2020PASP..132e4401Z}. Dramatically
increasing the number of telescopes available to students for
education, outreach, and public engagement makes it possible for young
people to access and participate in research, and can help to
encourage people from all backgrounds into a science
career\cite{Sadler2001,2011ASPC..443..162G,2017AstRv..13...28G}. BiSON:NG
offers an unprecedented opportunity for multifaceted science,
engagement, and collaboration.

%%%%%%%%%%%%%%%%%%%%%%%%%%%%%%%%%%%%%%%%%%%%%%%%%%%%%%%%%%%%%%%%%%%%%%%%%%%%%%%%

%%%%%%%%%%%%%%%%%%%%%%%%%%%%%%%%%%%%%%%%%%%%%%%%%%%%%%%%%%%%%%%%%%%%%%%%%%%%%%%%

\acknowledgments % equivalent to \section*{ACKNOWLEDGMENTS}       

We would like to thank all those who have been associated with {BiSON}
over the years.  We particularly acknowledge the technical assistance
at our remote network sites, with sincere apologies to anyone
inadvertently missed:
At Mount Wilson: Ed\,J.\,{Rhodes},\,Jr., Stephen Pinkerton, the team of
{USC} undergraduate observing assistants, former {USC} staff members
Maynard Clark, Perry Rose, Natasha Johnson, Steve Padilla, and Shawn
Irish, and former {UCLA} staff members Larry Webster and John Boyden.
At Las Campanas: Patricio Pinto, Andres Fuentevilla, Emilio Cerda,
Frank Perez, Marc Hellebaut, Patricio Jones, Gast{\'o}n Gutierrez,
Juan Navarro, Francesco Di Mille, Roberto Bermudez, and the staff of
{LCO}.
At Iza{\~n}a: Pere {Pall{\'e}}, Teo {Roca Cort{\'e}s}, Antonio
Pimienta, and the team of operators who have contributed to running
the {Mark~I} instrument over many years.
At Sutherland: Pieter Fourie, Willie Koorts, Jaci Cloete, Reginald
Klein, John Stoffels, Brendt Christian, and the staff of {SAAO}.
At Carnarvon: Les Bateman, Les Schultz, Sabrina Dowling-Giudici, Inge
Lauw of Williams and Hughes Lawyers, and {NBN} Co.\ Ltd.
At Narrabri: Mike Hill and the staff of {CSIRO}.
The authors are grateful for the financial support of the Science and
Technology Facilities Council ({STFC}), grant reference ST/R000417/1.
Additional funding was secured via the {STFC} Impact Accelerator
account, with the assistance of Alan Tibbatts from University of
Birmingham Enterprise. Funding for the Stellar Astrophysics Centre
(SAC) is provided by The Danish National Research Foundation, grant
reference DNRF106.

%%%%%%%%%%%%%%%%%%%%%%%%%%%%%%%%%%%%%%%%%%%%%%%%%%%%%%%%%%%%%%%%%%%%%%%%%%%%%%%%

% References
\bibliography{spie} % bibliography data in references.bib
\bibliographystyle{spiebib} % makes bibtex use spiebib.bst

\end{document}